\documentclass[final
,numberedheadings 
  ]
  {aipproc}
\layoutstyle{6x9}

\def \bea{\begin{eqnarray}}
\def \beq{\begin{equation}}
\def \b{{\cal B}}

\def \cB{{\cal B}}
\def \eea{\end{eqnarray}}
\def \eeq{\end{equation}}
\def \efi{Enrico Fermi Institute Report No.\ }
\def \gb{\overline{\Gamma}}
\def \gs{\stackrel{>}{\sim}}
\def \ket#1{|#1 \rangle}
\def \ob{\overline{B}^0}
\def \od{\overline{D}^0}
\def \ol{\overline}
\def \s{\sqrt{2}}
\def \st{\sqrt{3}}

\begin{document}

\title{$B$ Physics (Theory)\footnote{\efi 03-32, hep-ph/0306284.  Presented at
Fourth Tropical Workshop, Cairns, Australia, 9--13 June 2003.  Proceedings
to be published by AIP.}}

\author{Jonathan L. Rosner}{
  address={Enrico Fermi Institute and Department of Physics\\
University of Chicago, 5640 S. Ellis Avenue, Chicago IL 60637}
}

\begin{abstract}
Some theoretical aspects of $B$ physics are reviewed.  These include a brief
recapitulation of information on weak quark transitions as described by the
Cabibbo-Kobayashi-Maskawa (CKM) matrix, descriptions of CP asymmetries in
$B$ decays to CP eigenstates and to self-tagging modes, a discussion of
final-state phases in $B$ and charm decays, some topics on $B_s$ properties
and decays, prospects for unusual excited $B$ states opened by discovery
of some narrow $c \bar s$ resonances, and the search for heavier
$Q=1/3$ quarks predicted in some extended grand unified theories.
\end{abstract}

\maketitle

\section{Introduction}

The physics of $B$ mesons (those containing the $b$ [bottom or beauty] quark)
has greatly illuminated the study of the electroweak and strong interactions.
This brief review is devoted to some theoretical aspects of $B$ physics, with
emphasis on current questions for $e^+ e^-$ and hadron collider experiments.
Section 2 reviews weak quark transitions.  We note in Section 3 progress and
puzzles in the study of $B^0$ decays to CP eigenstates, turning in Section 4
to direct CP asymmetries which require strong final-state phases for their
observation.  Some aspects of these phases are described in in Section 5.  We
devote Section 6 to the strange $B$ mesons, with Section 7 treating the
possibility of narrow $b \bar s$ states suggested by the recent observation of
narrow $c \bar s$ mesons.  Section 8 discusses the prospects for seeing heavier
$Q=1/3$ quarks. We summarize in Section 9.  This review updates and
supplements Refs.\ \cite{JRmor,JRLHC}.
 
\section{Weak quark transitions}
\begin{figure}
\includegraphics[height=3in]{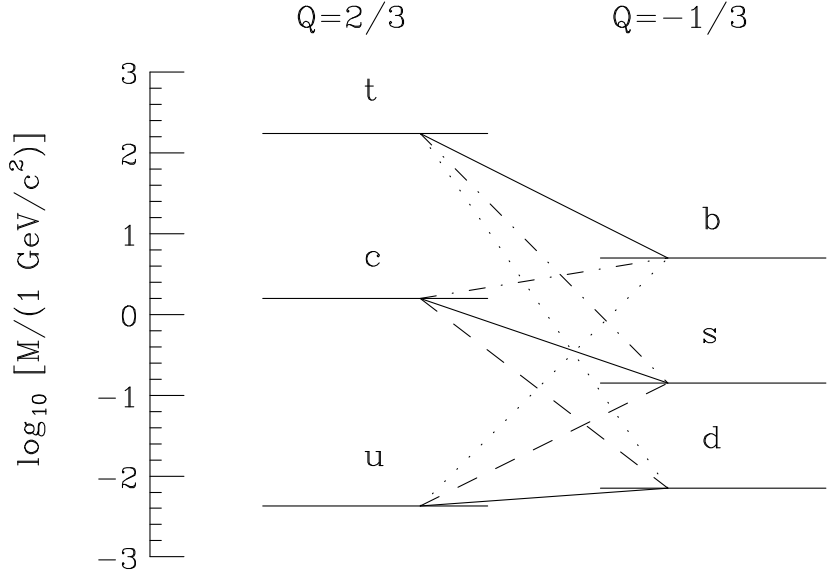}
\caption{Charge-changing weak transitions among quarks.  Solid
lines:  relative strength 1; dashed lines:  relative strength 0.22;
dot-dashed lines:  relative strength 0.04; dotted lines:  relative strength
$\le 0.01$.
\label{fig:trans}}
\end{figure}

The relative strengths of charge-changing weak quark transitions are shown
in Fig.\ \ref{fig:trans}.  It is crucial to describe this pattern precisely in
order to distinguish among theories which might predict it, and to see whether
it can reproduce all weak phenomena including CP violation or whether some new
ingredient is needed.

\subsection{The CKM matrix}
The interactions in Fig.\ \ref{fig:trans} may be parametrized by a unitary
Cabibbo-Kobayashi-Maskawa (CKM) matrix which can be written approximately
\cite{WP,Battaglia} in terms of a small expansion parameter $\lambda$ as
\beq \label{eqn:WP}
V_{\rm CKM} = \left[ \begin{array}{c c c}
1 - \frac{\lambda^2}{2} & \lambda & A \lambda^3 (\rho - i \eta) \\
- \lambda & 1 - \frac{\lambda^2}{2} & A \lambda^2 \\
A \lambda^3 (1 - \bar \rho - i \bar \eta) & - A \lambda^2 & 1 \end{array}
\right]~~~,
\eeq
where $\bar \rho \equiv \rho (1 - \frac{\lambda^2}{2})$ and
$\bar \eta \equiv \eta (1 - \frac{\lambda^2}{2})$.
The columns refer to $d,s,b$ and the rows to $u,c,t$.  The parameter $\lambda =
0.224$ \cite{Battaglia} is $\sin \theta_c$, where $\theta_c$ is the Cabibbo
angle.  The value $|V_{cb}| \simeq 0.041$, obtained from $b \to c$ decays,
indicates $A \simeq 0.82$, while $|V_{ub}/V_{cb}| \simeq 0.1$, obtained from
$b \to u$ decays, implies $(\rho^2 + \eta^2)^{1/2} \simeq 0.45$.  We shall
generally use the CKM parameters quoted in Ref.\ \cite{CKMf}.

\subsection{The unitarity triangle}

\begin{figure}
\includegraphics[height=1.5in]{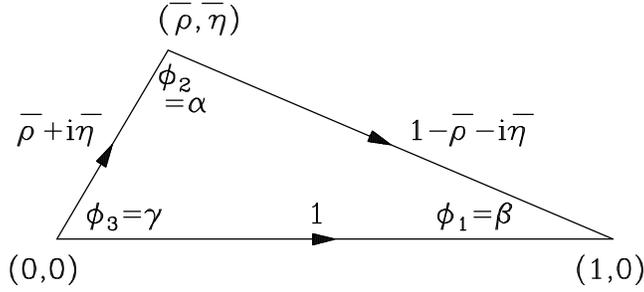}
\caption{The unitarity triangle \cite{JRmor}.  Ranges of angles allowed at
95\% c.l.\ \cite{CKMf} are $78^\circ < \alpha < 122^\circ$, $20^\circ < \beta
< 27^\circ$, and $38^\circ < \gamma < 80^\circ$.
\label{fig:ut}}
\end{figure}

The unitarity of the CKM matrix can be expressed in terms of a triangle in the
complex $\bar \rho + i \bar \eta$ plane, with vertices at (0,0) (angle $\phi_3
= \gamma$), (1,0) (angle $\phi_1 = \beta$), and $(\bar \rho, \bar \eta)$ (angle
$\phi_2 = \alpha$).  The triangle has unit base and its other two sides are
$\bar \rho + i \bar \eta = -(V^*_{ub}V_{ud}/ V^*_{cb} V_{cd})$
$\phi_1 = \beta$) and $1 - \bar \rho - i \bar \eta = -(V^*_{tb}V_{td}/V^*_{cb}
V_{cd})$.  The result is shown in Fig.\ \ref{fig:ut}.

\begin{figure}
\includegraphics[height=5.9in]{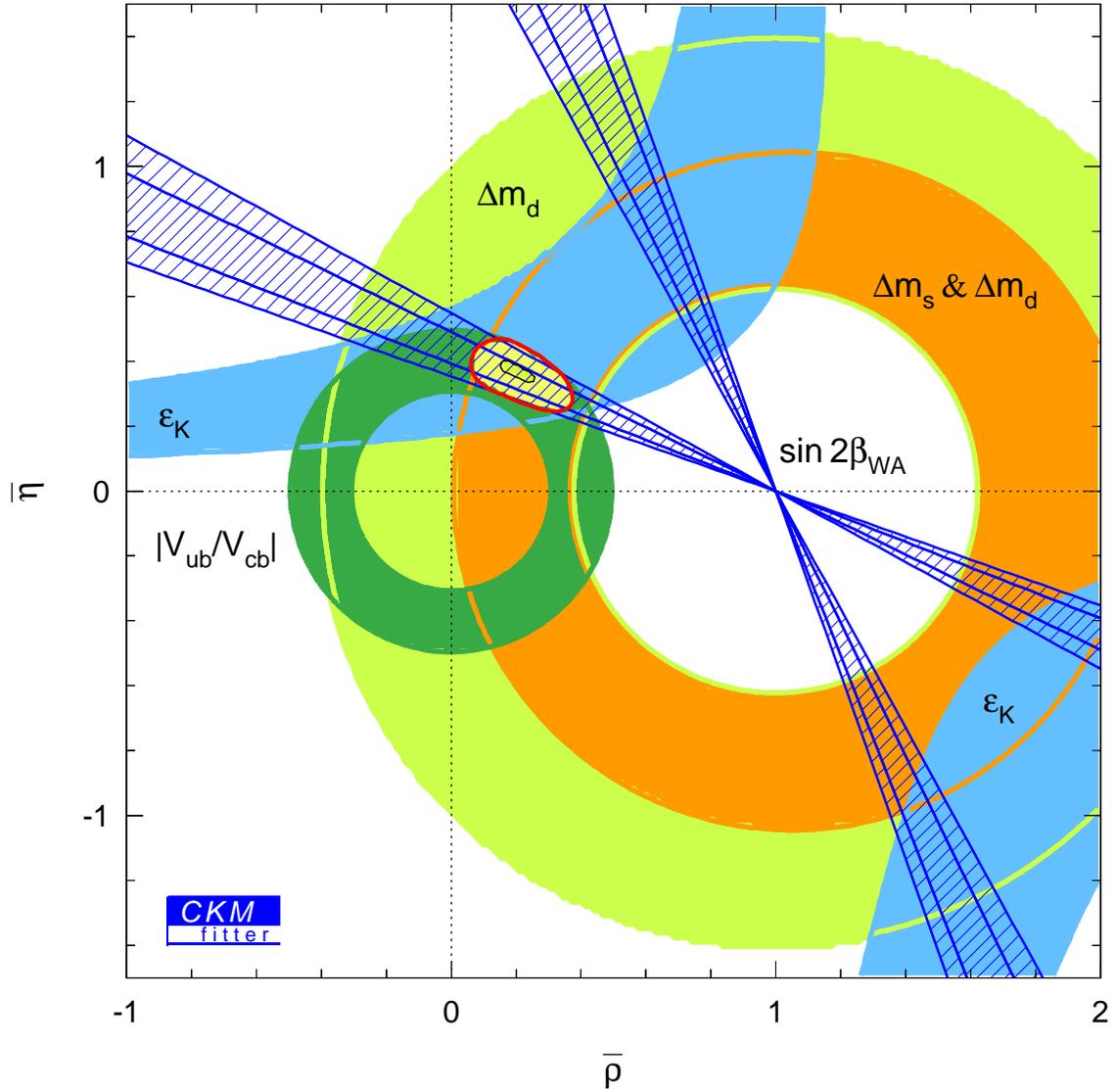}
\caption{Constraints in the $(\bar \rho, \bar \eta)$ plane as of July
2002 (from the web page of Ref.\ \cite{CKMf}).
\label{fig:CKMf}}
\end{figure}

Flavor-changing loop diagrams provide further constraints.  CP-violating
$K^0$--$\ol K^0$ mixing is dominated by $\bar s d \to \bar d s$ with virtual
$t \bar t$ and $W^+ W^-$ intermediate states.  It constrains Im$(V_{td}^2) \sim
\bar \eta (1 - \bar \rho)$, giving a hyperbolic band in the $(\bar \rho, \bar
\eta)$ plane.  $B^0$--$\ol B^0$ mixing is dominated by $t \bar t$ and $W^+ W^-$
in the loop diagram for $\bar b d \to \bar d b$, and thus constrains $|V_{td}|$
and hence $|1 - \bar \rho - i \bar \eta|$.  By comparing $B_s$--$\ol B_s$ and
$B^0$--$\ol B^0$ mixing, one reduces dependence on matrix elements and
learns $|V_{ts}/V_{td}| > 4.4$ or $|1 - \bar \rho - i \bar \eta| < 1$. 
The resulting constraints are shown in Fig.\ \ref{fig:CKMf} \cite{CKMf}.

\section{$B$ decays to CP eigenstates}
The decays of neutral $B$ mesons to CP eigenstates $f$, where $CP \ket{f} =
\xi_f \ket{f}$, $\xi_f = \pm 1$, provide direct information on CKM phases
without the need to understand complications of strong interactions.  As a
result of $B^0$--$\ob$ mixing, a state $B^0$ at proper time $t=0$ evolves into
a mixture of $B^0$ and $\ob$ denoted $B^0(t)$.  Thus there will be one pathway
to the final state $f$ from $B^0$ through the amplitude $A$ and another from
$\ob$ through the amplitude $\bar A$, which acquires an additional phase $2
\phi_1 = 2 \beta$ through the mixing.  The interference of these two amplitudes
can differ in the decays $B^0(t) \to f$ and $\ob (t) \to f$, leading to a
time-integrated rate asymmetry
\beq
A_{CP} \equiv \frac{\Gamma(\ob \to f) - \Gamma(B^0 \to f)}
                          {\Gamma(\ob \to f) + \Gamma(B^0 \to f)}
\eeq
as well as to time-dependent rates
\beq
\left\{ \begin{array}{c} \Gamma[B^0(t) \to f] \\ \Gamma[\ob (t) \to f]
\end{array} \right\} \sim e^{- \Gamma t} [ 1 \mp A_f \cos \Delta m t
 \mp S_f \sin \Delta m t ]~~~,
\eeq
where
\beq
A_f \equiv \frac{|\lambda|^2 - 1}{|\lambda|^2 + 1}~~,~~~
S_f \equiv \frac{2 {\rm Im} \lambda}{|\lambda|^2 + 1}~~,~~~
\lambda \equiv e^{-2 i \beta} \frac{\bar A}{A}~~~,
\eeq
where $S_f^2 + A_f^2 \le 1$ \cite{BaBarPhys,TASI}.

\subsection{$B^0 \to J/\psi K_S$ and $\phi_1 = \beta$}
For this decay one has $\bar A/A \simeq \xi_{J/\psi K_S} = -1$.  The
time-integrated asymmetry $A_{CP}$ is proportional to $\sin(2
\phi_1) = \sin(2 \beta)$.  Using this and related decays involving the same
quark subprocess, BaBar \cite{Babeta} finds $\sin(2 \beta) = 0.741 \pm 0.067
\pm 0.033$ while Belle \cite{Bebeta} finds $0.719 \pm 0.074 \pm 0.035$.  The
world average \cite{avbeta} is $\sin(2 \beta) = 0.734 \pm 0.054$, consistent
with other determinations \cite{CKMf,Ciu,AL}.

\subsection{$B^0 \to \pi^+ \pi^-$ and $\phi_2 = \alpha$}
Two amplitudes contribute to the decay:  a ``tree'' $T$ and a ``penguin'' $P$:
\beq
A = - (|T|e^{i \gamma} + |P| e^{i \delta})~~,~~~
\bar A = - (|T|e^{-i \gamma} + |P| e^{i \delta})~~~,
\eeq
where $\delta$ is the relative $P/T$ strong phase.  The asymmetry
$A_{CP}$ would be proportional to $\sin(2 \alpha)$ if the penguin
amplitude could be neglected.  One way to account for its contribution is via
an isospin analysis \cite{GL} of $B$ decays to $\pi^+ \pi^-$, $\pi^\pm \pi^0$,
and $\pi^0 \pi^0$, separating the amplitudes for decays involving $I=0$ and
$I=2$ final states.  Information can then be obtained on both strong and weak
phases.  Since the branching ratio of $B^0$ to $\pi^0 \pi^0$ may be very small,
of order $10^{-6}$, alternative methods \cite{GR02,GRconv} may be useful in
which flavor SU(3) symmetry is used to estimate the penguin contribution
\cite{SW,GHLR,Charles}.

The tree amplitude for $B^0 (= \bar b d) \to \pi^+ \pi^-$ involves
$\bar b \to \pi^+ \bar u$, with the spectator $d$ quark combining
with $\bar u$ to form a $\pi^-$.  Its magnitude is $|T|$; its weak phase
is Arg($V^*_{ub}) = \gamma$; by convention its strong phase is 0.  The
penguin amplitude involves the flavor structure $\bar b \to \bar d$, with the
final $\bar d d$ pair fragmenting into $\pi^+ \pi^-$.  Its magnitude is
$|P|$.  The dominant $t$ contribution in the loop diagram for $\bar b \to \bar
d$ can be integrated out and the unitarity relation $V_{td} V^*_{tb} =
- V_{cd} V^*_{cb} - V_{ud} V^*_{ub}$ used.  The $V_{ud} V^*_{ub}$ contribution
can be absorbed into a redefinition of the tree amplitude, after which
the weak phase of the penguin amplitude is 0 (mod $\pi$).  By definition, its
strong phase is $\delta$.

The time-dependent asymmetries $S_{\pi \pi}$ and $A_{\pi \pi}$
specify both $\gamma$ (or $\alpha = \pi - \beta - \gamma$) and $\delta$,
if one has an independent estimate of $|P/T|$.  One may obtain $|P|$ from $B^+
\to K^0 \pi^+$ using flavor SU(3) \cite{SW,GHLR,GR95} and $|T|$ from $B \to
\pi l \nu$ using factorization \cite{LR}.  (An alternative method discussed
in Refs.\ \cite{GRconv,Charles} uses the measured ratio of the $B^+ \to K^0
\pi^+$ and $B^0 \to \pi^+ \pi^-$ branching ratios to constrain $|P/T|$.)

\begin{figure}
\includegraphics[height=4.5in]{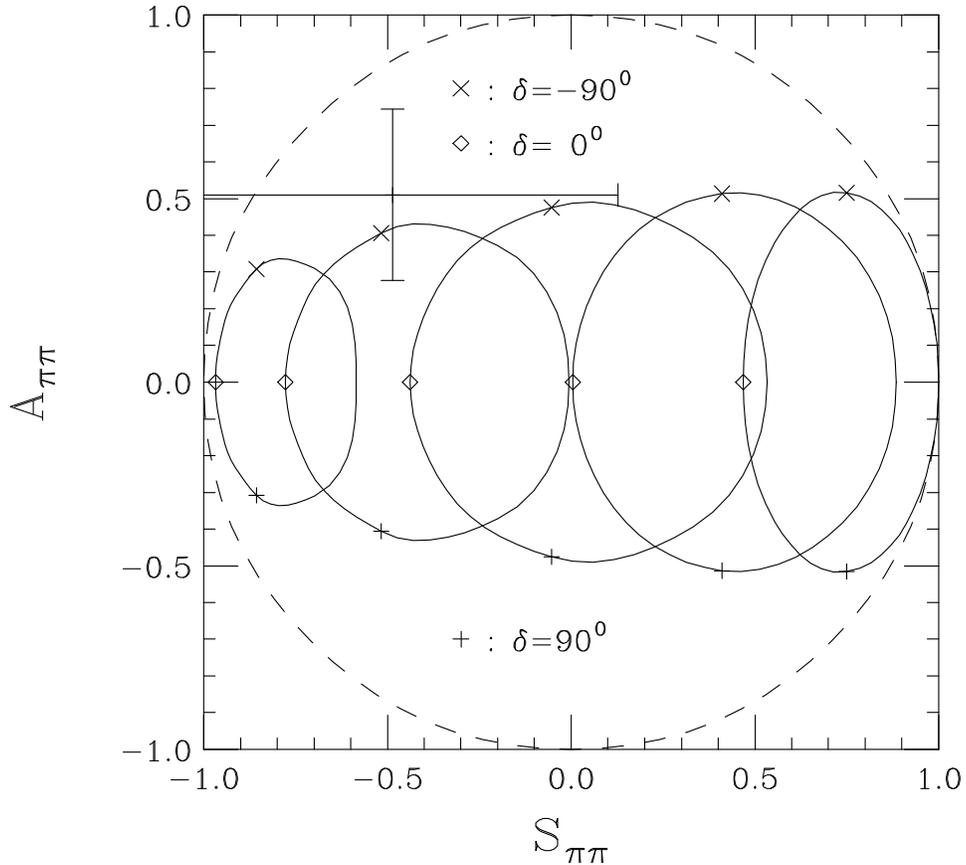}
\caption{Curves depicting dependence of $S_{\pi \pi}$ and
$A_{\pi \pi}$ on $\delta$ \cite{JRmor}.
From right to left the curves correspond to $\phi_2 = (120^\circ, 105^\circ,
90^\circ, 75^\circ, 60^\circ)$.  Plotted point:  average of BaBar and Belle
values (see text).  As $|\delta|$ increases from 0 to $\pi$, the values of
$S_{\pi \pi}$ become more positive, while the magnitudes $|A_{\pi
\pi}|$ increase from zero and then return to zero.  Positive values of
$A_{\pi \pi}$ correspond to negative values of $\delta$.
\label{fig:sa}}
\end{figure}

In addition to $S_{\pi \pi}$ and $A_{\pi \pi}$, a useful quantity
is the ratio of the $B^0 \to \pi^+ \pi^-$ branching ratio $\cB(\pi^+ \pi^-)$
(unless otherwise specified, branching ratios refer to CP averages) to that due
to the tree amplitude alone:
\beq
R_{\pi \pi} \equiv \frac{\cB(\pi^+ \pi^-)}{\cB(\pi^+ \pi^-)|_{\rm tree}}
= 1 + 2 \left| \frac{P}{T} \right| \cos \delta \cos \gamma
+ \left| \frac{P}{T} \right|^2~~~.
\eeq
One also has
\beq
R_{\pi \pi} S_{\pi \pi} = \sin 2 \alpha + 2 \left| \frac{P}{T} \right|
\cos \delta \sin(\beta - \alpha) - \left| \frac{P}{T} \right|^2 \sin 2 \beta
~~~,
\eeq
\beq
R_{\pi \pi} A_{\pi \pi} = - 2 |P/T| \sin \delta \sin \gamma
~~~.
\eeq
We take $\beta = 23.6^\circ$.
The value of $|P/T|$ (updating \cite{GR02,GRconv}) is $0.28 \pm 0.06$.  Taking
the central value, we plot in Fig.\ \ref{fig:sa} trajectories in the
($S_{\pi \pi},A_{\pi \pi}$) plane for $-\pi \le \delta \le \pi$.

\begin{table}
\caption{Values of $S_{\pi \pi}$ and $A_{\pi \pi}$ quoted by
BaBar and Belle and their averages.  Here we have applied scale factors
of $\sqrt{\chi^2} = (2.31,1.24)$ to the errors for
$S_{\pi \pi}$ and $A_{\pi \pi}$, respectively.
\label{tab:sa}}
\begin{tabular}{c c c c} \hline
    Quantity         & BaBar \cite{Bapipi}  & Belle \cite{Bepipi}            &
    Average \\ \hline
$S_{\pi \pi}$ & $0.02\pm0.34\pm0.05$ & $-1.23\pm0.41^{+0.08}_{-0.07}$ &
    $-0.49 \pm 0.61$ \\
$A_{\pi \pi}$ & $0.30\pm0.25\pm0.04$ & $ 0.77 \pm 0.27 \pm 0.08$ &
    $~~0.51 \pm 0.23$ \\ \hline
\end{tabular}
\end{table}

\begin{figure}
\includegraphics[height=3.1in]{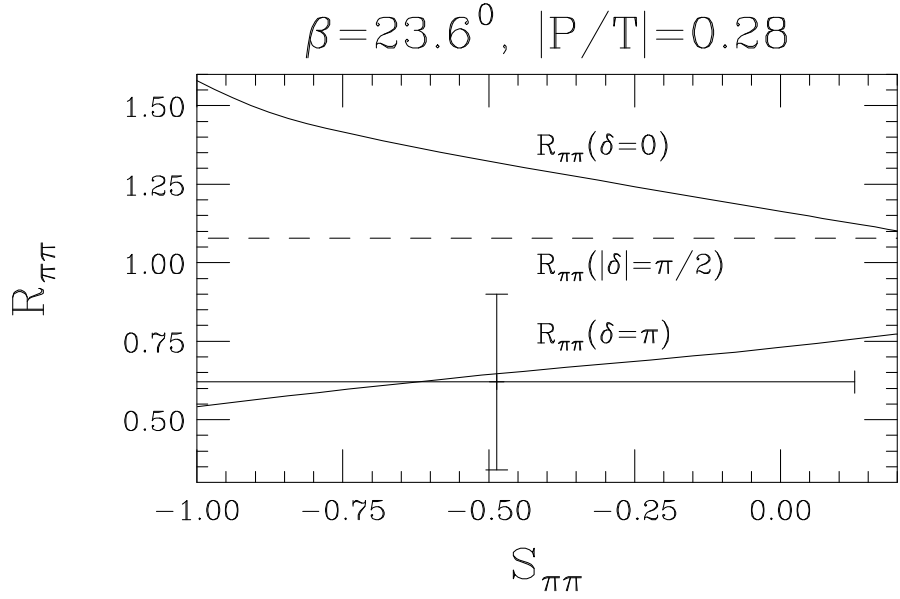}
\caption{Curves depicting dependence of $R_{\pi \pi}$ on $S_{\pi \pi}$
for various values of $\delta$\cite{JRmor}.  The plotted point is the average
of BaBar and Belle values for $ S_{\pi \pi}$ (see text).
\label{fig:sr}}
\end{figure}

As shown in Table \ref{tab:sa}, BaBar \cite{Bapipi} and Belle \cite{Bepipi}
obtain different asymmetries, especially $S_{\pi \pi}$.  Even once this
conflict is resolved, there are discrete ambiguities, since curves for
different $\alpha$ intersect one another.  These can be resolved with the
help of $R_{\pi \pi} = 0.62 \pm 0.28$, as shown in Fig.\ \ref{fig:sr}.  The
present value favors large $|\delta|$ and $\phi_2 = \alpha > 90^\circ$, but
with large uncertainty.  It is not yet settled whether $A_{\pi \pi} \ne 0$,
corresponding to ``direct'' CP violation.

\begin{figure}
\includegraphics[height=3.8in]{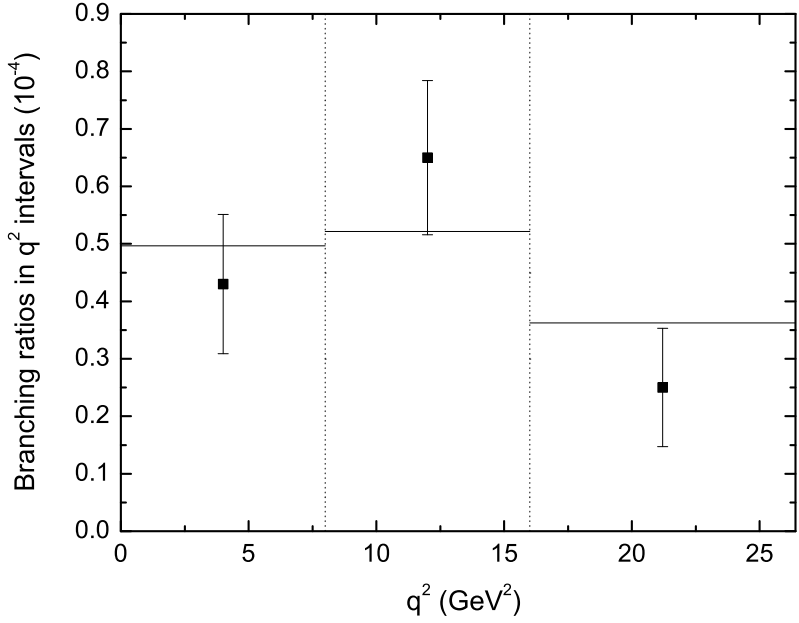}
\caption{Fit to $\int dq^2 \frac{d \b}{d q^2}(B^0 \to \pi^- l^+ \nu_l)$ values
\cite{LR03} obtained for three $q^2$ bins in Ref.\ \cite{CLEOsl03}.  Points
with errors correspond to data; the histogram represents the fit.
\label{fig:br}}
\end{figure}

\begin{figure}
\includegraphics[height=5.0in]{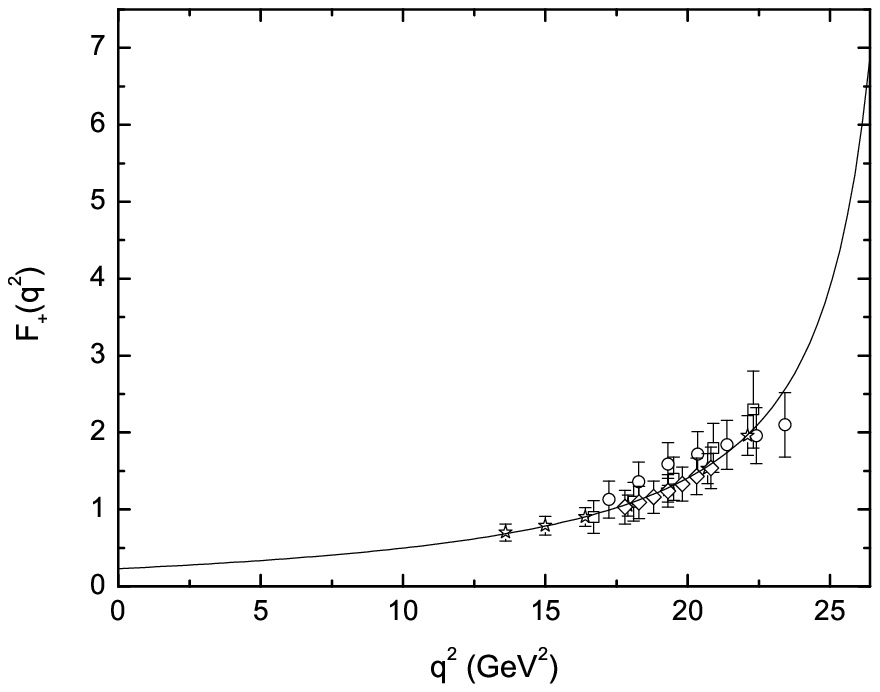} \\
\caption{Comparison of lattice data points with best-fit form factor $F_+(q^2)$
\cite{LR03}. Lattice data are from UKQCD (squares), APE
(stars), Fermilab (circles) and JLQCD (diamonds) (see \cite{LR03}).
\label{fig:ff}}
\end{figure}

Does the tree ($T$) amplitude alone account for the $B^0 \to \pi^+ \pi^-$ rate
(corresponding to $R_{\pi \pi} = 1$) or is there destructive
interference with the penguin terms (corresponding to $R_{\pi \pi} < 1$)?
Recently Zumin Luo and I \cite{LR03} have combined the $B \to \pi l \nu$
spectrum reported by the CLEO Collaboration \cite{CLEOsl03} with information
on the $B^+ \to \pi^+ \pi^0$ rate, estimates of the ratio of color-suppressed
to color-favored amplitude in this process, other determinations of $|V_{ub}|$,
and lattice gauge theory predictions of the $B \to \pi l \nu$ form factor at
high momentum transfer, to find that $R_{\pi \pi} = 0.87^{+0.11}_{-0.28}$.
The corresponding fit to the $B \to \pi l \nu$ spectrum is shown in Fig.\
\ref{fig:br}, while the fit to the lattice predictions is shown in Fig.\
\ref{fig:ff}.  For massless leptons (a good approximation), the differential
decay rate is governed by a single form factor $F_+(q^2)$:
\beq \label{eqn:diff}
\frac{d\Gamma}{dq^2}(B^0 \to \pi^-\ell^+ \nu_{\ell}) =
\frac{G_F^2|V_{ub}|^2}{24\pi^3}|\vec{p}_{\pi}|^3|F_+(q^2)|^2~~ ,
\eeq
where we take the simple form $F_+(q^2) = [F(0)](1 + a q^2/m_{B^*}^2)/(1 -
q^2/m_{B^*}^2)$.  We find $a = 1.14^{+0.72}_{-0.42}$, $F_+(0) = 0.23 \pm 0.04$.
The evidence for destructive tree-penguin interference in $B^0 \to \pi^+
\pi^-$ is not overwhelming.  A more definite conclusion will be possible when
improved $B \to \pi l \nu$ spectra become available.

\subsection{$B^0 \to \phi K_S$ vs. $B^0 \to J/\psi K_S$}
In $B^0 \to \phi K_S$, governed by the $\bar b \to \bar s$ penguin amplitude,
the standard model predicts the same CP asymmetries as in those processes (like
$B^0 \to J/\psi K_S$)  governed by $\bar b \to \bar s c \bar c$.  In both cases
the weak phase is expected to be 0 (mod $\pi$), so the indirect CP asymmetry
should be governed by $B^0$--$\ob$ mixing and thus should be proportional to
$\sin 2 \beta$.  There should be no direct CP asymmetries (i.e., $A
\simeq 0$) in either case.  This is true for $B \to J/\psi K$; $A$ is
consistent with zero in the neutral mode, while the direct CP asymmetry is
consistent with zero in the charged mode \cite{Babeta}.  However, a different
result for $B^0 \to \phi K_S$ could point to new physics in the $\bar b \to
\bar s$ penguin amplitude \cite{GW}.

The experimental asymmetries in $B^0 \to \phi K_S$ \cite{Baphks,Bephks} are
shown in Table \ref{tab:phks}. For $A_{\phi K_S}$ there is a substantial
discrepancy between BaBar and Belle.  The value of $S_{\phi K_S}$, which
should equal $\sin 2 \beta = 0.734 \pm 0.054$ in the standard model, is about
$2.7 \sigma$ away from it.  If the amplitudes for $B^0 \to \phi K^0$ and $B^+
\to \phi K^+$ are equal (true in many approaches), the time-integrated CP
asymmetry $A_{CP}$ in the charged mode should equal $A_{\phi K_S}$.  The
BaBar Collaboration \cite{Aubert:2003tk} has recently reported
$A_{CP}(\phi K^+) = 0.039 \pm 0.086 \pm 0.011$.

\begin{table}
\caption{Values of $S_{\phi K_S}$ and $A_{\phi K_S}$ quoted by
BaBar and Belle and their averages.  We have applied a scale factor of
$\sqrt{\chi^2} = 2.29$ to the error on $A_{\phi K_S}$.
\label{tab:phks}}
\begin{tabular}{c c c c} \hline
    Quantity         & BaBar \cite{Baphks}  & Belle \cite{Bephks}            &
    Average \\ \hline
$S_{\phi K_S}$ & $-0.18\pm0.51\pm0.07$ & $-0.73\pm0.64\pm0.22$ &
 $-0.38 \pm 0.41$ \\
$A_{\phi K_S}$ & $0.80\pm0.38\pm0.12$ & $-0.56\pm0.41\pm0.16$ &
 $0.19 \pm 0.68$ \\ \hline
\end{tabular}
\end{table}

Many proposals for new physics can account for the departure of $S_
{\phi K_S}$ from its expected value of $\sin 2 \beta$ \cite{npphks}.  A method
for extracting a new physics amplitude has been developed \cite{CR03}, using
the measured values of $S_{\phi K_S}$ and $A_{\phi K_S}$ and the ratio
\beq \label{eqn:rphks}
R_{\phi K_S} \equiv \frac{\cB(B^0 \to \phi K_S)}{\cB(B^0 \to \phi K_S)|_{\rm
 std}} = 1 + 2 r \cos \phi \cos \delta + r^2~~~,
\eeq
where $r$ is the ratio of the magnitude of the new amplitude to the one in
the standard model, and $\phi$ and $\delta$ are their relative weak and
strong phases.  For any values of $R_{\phi K_S}$, $\phi$, and $\delta$, Eq.\
(\ref{eqn:rphks}) can be solved for the amplitude ratio $r$ and one then
calculates the asymmetry parameters as functions of $\phi$ and $\delta$.
The $\phi K_S$ branching ratio in the standard model is calculated using the
penguin amplitude from $B^+ \to K^{*0} \pi^+$ and an estimate of electroweak
penguin corrections.  Various regions of $(\phi, \delta)$ can reproduce the
observed values of $S_{\phi K_S}$ and $A_{\phi K_S}$.  Typical
values of $r$ are of order 1; one generally needs to invoke new-physics
amplitudes comparable to those in the standard model.

The above scenario envisions new physics entirely in $B^0 \to \phi K^0$ and
not in $B^+ \to K^{*0} \pi^+$.  An alternative is that new physics
contributes to the $\bar b \to \bar s$ penguin amplitude and thus appears
in {\it both} decays.  Again, $S_{\phi K_S}$ suggests
an amplitude associated with new physics \cite{CR03}, but one must wait until
the discrepancy with the standard model becomes more significant.  At present
both the decays $B^0 \to K_S (K^+ K^-)_{CP = +}$ and $B^0 \to \eta' K_S$
display CP asymmetries consistent with standard expectations.

\subsection{$B^0 \to K_S (K^+ K^-)_{CP=+}$}
The Belle Collaboration \cite{Bephks} finds that for $K^+ K^-$ not in the
$\phi$ peak, most of the decay $B^0 \to K_S K^+ K^-$ involves even CP for the
$K^+ K^-$ system ($\xi_{K^+ K^-} = +1$).  It is found that
\bea
- \xi_{K^+ K^-} S_{K^+ K^-} & = & 0.49\pm0.43\pm 0.11^{+0.33}_{-0.00}
~~~,\\
A_{K^+ K^-} & = & -0.40 \pm 0.33 \pm 0.10^{+0.00}_{-0.26}~~,
\eea
where the third set of errors arise from uncertainty in the fraction of the
CP-odd component.  Independent estimates of this fraction have been performed
in Refs.\ \cite{GLNQ} and \cite{GRKKK}.  The quantity $- \xi_{K^+ K^-}
S_{K^+ K^-}$ should equal $\sin 2 \beta$ in the standard model, but additional
non-penguin contributions can lead this quantity to range between 0.2 and
1.0 \cite{GRKKK}.

\subsection{$B \to \eta' K$ (charged and neutral modes)}
At present neither the rate nor the CP asymmetry in $B \to \eta' K$ present
a significant challenge to the standard model.  The rate can be reproduced
with the help of a modest contribution from a ``flavor-singlet penguin''
amplitude \cite{DGR95,DGR97,CR01,FHH,CGR}.  (An alternative treatment
\cite{BN} finds an enhanced standard-penguin contribution to $B \to \eta' K$.)
The CP asymmetry is not a problem; the ordinary and singlet penguin amplitudes
have the same weak phase Arg$(V^*_{ts}V_{tb}) \simeq \pi$ and hence one
expects $S_{\eta' K_S} \simeq \sin 2 \beta$, $A_{\eta' K_S}
\simeq 0$.  The experimental situation is shown in Table \ref{tab:etapks}.
The value of $S_{\eta' K_S}$ is consistent with the standard model
expectation at the $1 \sigma$ level, while $A_{\eta' K_S}$ is consistent
with zero.

\begin{table}
\caption{Values of $S_{\eta' K_S}$ and $A_{\eta' K_S}$ quoted by
BaBar and Belle and their averages.  We have applied scale factors
$\sqrt{\chi^2} = (1.48,1.15)$ to the errors for
$S_{\eta' K_S}$ and $A_{\eta' K_S}$, respectively.
\label{tab:etapks}}
\begin{tabular}{c c c c} \hline
    Quantity         & BaBar \cite{Baphks}  & Belle \cite{Bephks}            &
    Average \\ \hline
$S_{\eta' K_S}$ & $0.02\pm0.34\pm0.03$ & $0.76\pm0.36^{+0.05}_{-0.06}$ &
$0.37 \pm 0.37$  \\
$A_{\eta' K_S}$ & $-0.10\pm0.22\pm0.03$ & $0.26\pm0.22\pm0.03$ &
$0.08 \pm 0.18$  \\ \hline
\end{tabular}
\end{table}

The singlet penguin amplitude may contribute elsewhere in $B$ decays.  It is
a possible source of a low-effective-mass $\bar p p$ enhancement \cite{Kpp} in
$B^+ \to \bar p p K^+$ \cite{JRbbbar}.

\section{Direct CP asymmetries}
Decays such as $B \to K \pi$ (with the exception of $B^0 \to K^0 \pi^0$) are
{\it self-tagging}:  Their final states indicate the flavor of the
decaying state.  For example, the $K^+ \pi^-$ final state is expected to
originate purely from a $B^0$ and not from a $\ob$.  Such self-tagging decays
involve both weak and strong
phases.  Several methods permit one to separate these from one another.

\subsection{$B^0 \to K^+ \pi^-$ vs.\ $B^+ \to K^0 \pi^+$}
The decay $B^+ \to K^0 \pi^+$ is a pure penguin ($P$) process, while the
amplitude for $B^0 \to K^+ \pi^-$ is proportional to $P + T$, where $T$ is a
(strangeness-changing) tree amplitude.  The ratio $T/P$ has magnitude $r$, weak
phase $\gamma \pm \pi$, and strong phase $\delta$.  The ratio
$R_0$ of these two rates (averaged over a process and its CP conjugate) is
\beq \label{eqn:Rval}
R_0 \equiv \frac{\gb(B^0 \to K^+ \pi^-)}{\gb(B^+ \to K^0 \pi^+)} =
1 - 2 r \cos \gamma \cos \delta + r^2 \ge \sin^2 \gamma~~~,
\eeq
where the inequality holds for any $r$ and $\delta$.  For $R_0 < 1$ this
inequality implies a constraint on $\gamma$ \cite{FM}.  Using branching ratios
\cite{Babrs,Bebrs,CLbrs} averaged in Ref.\ \cite{GRKpi03} and the $B^+/B^0$
lifetime ratio from Ref.\ \cite{LEPBOSC}, one finds $R_0 = 0.948 \pm 0.074$,
which is consistent with 1 and does not permit application of the bound.
However, using additional information on $r$ and the CP asymmetry in $B^0 \to
K^+ \pi^-$, one can obtain a constraint on $\gamma$ \cite{GR02,GRKpi}.

In Refs.\ \cite{GR02,GRKpi} we defined a ``pseudo-asymmetry'' normalized by the
rate for $B^0 \to K^0 \pi^+$, a process which should not have a CP asymmetry
since only the penguin amplitude contributes to it:
\beq \label{eqn:asy}
A_0 \equiv \frac{\Gamma(\ob \to K^- \pi^+) - \Gamma(B^0 \to K^+ \pi^-)}
{2 \gb(B^+ \to K^0 \pi^+)} = R_0 A_{CP}(K^+ \pi^-)
= - 2 r \sin \gamma \sin \delta~~~.
\eeq
One can eliminate $\delta$ between this equation and Eq.\ (\ref{eqn:Rval}) and
plot $R_0$ as a function of $\gamma$ for the allowed range of $|A_0|$.  For
a recent analysis based on this method see \cite{JRmor}.  Instead we shall
directly use $A_{CP}(K^+ \pi^-)$, as in Refs.\ \cite{JRLHC} and \cite{MGFPCP}.

\begin{figure}
\includegraphics[height=3.5in]{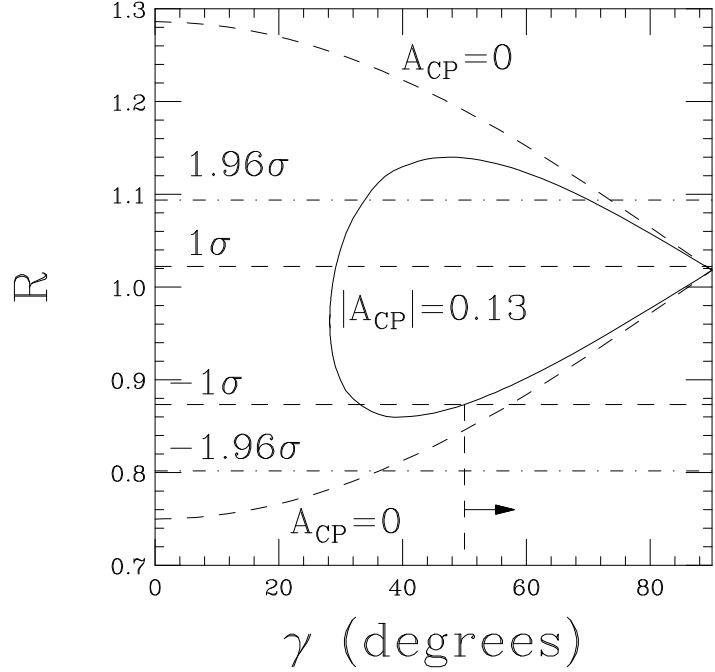}
\caption{Behavior of $R_0$ for $r = 0.134$ and $A_0 = 0$ (dashed curves) or
$|A_0| = 0.13$ (solid curve) as a function of the weak phase $\gamma$
\cite{GRKpi03}.
Horizontal dashed lines denote $\pm 1 \sigma$ experimental limits on $R_0$,
while dot-dashed lines denote $95\%$ c.l. ($\pm 1.96 \sigma$) limits.
\label{fig:Racp}}
\end{figure}

The value of $r$, based on present branching ratios and arguments given in
Refs.\ \cite{JRmor,GR02,GRKpi}) is $r = 0.17 \pm 0.04$.  BaBar and Belle data
imply $A_{CP}(K^+ \pi^-) = -0.09 \pm 0.04$, leading us to take its magnitude as
less than 0.13 at the $1 \sigma$ level.  Curves for $A_{CP}(K^+ \pi^-)=0$ and
$|A_{CP}(K^+ \pi^-)| = 0.13$ are shown in Fig.\ \ref{fig:Racp} \cite{GRKpi03}.
The lower limit $r = 0.13$ is used to generate these curves since the limit
on $\gamma$ will be the most conservative.

Using the $1 \sigma$ constraints on $R_0$ and $|A_{CP}(K^+ \pi^-)|$ one finds
$\gamma \gs 50^\circ$.  No bound can be obtained at the 95\% confidence level,
however.  Further data are needed in order for a useful constraint to be
obtained.

\subsection{$B^+ \to K^+ \pi^0$ vs.\ $B^+ \to K^0 \pi^+$}
The comparison of rates for $B^+ \to K^+ \pi^0$ and $B^+ \to K^0 \pi^+$ also
gives information on $\gamma$.  The amplitude for $B^+ \to K^+ \pi^0$ is
proportional to $P + T + C$, where $C$ is a color-suppressed amplitude.  It was
suggested in \cite{GRL} that this amplitude be compared with $P$ from $B^+ \to
K^0 \pi^+$ and $T+C$ taken from $B^+ \to \pi^+ \pi^0$ using flavor SU(3) and
a triangle construction to determine $\gamma$.  Electroweak penguin amplitudes
contributing in the $T+C$ term \cite{EWP} may be taken into account \cite{NR}
by noting that since $T+C$ corresponds to isospin $I(K \pi) = 3/2$ for the
final state, the strong-interaction phase of its EWP contribution is the same
as that of the rest of the $T+C$ amplitude.

\begin{figure}
\includegraphics[height=3.5in]{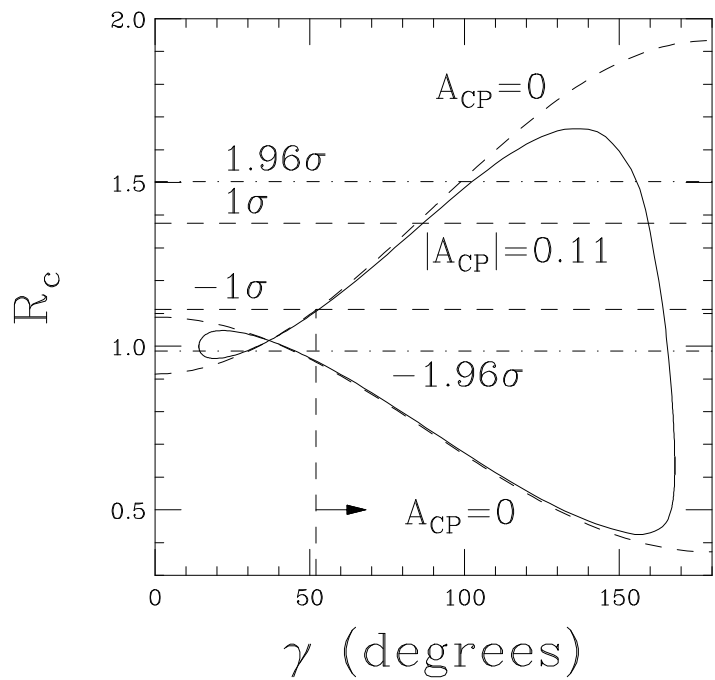}
\caption{Behavior of $R_c$ for $r_c = 0.22$ ($1 \sigma$ upper limit) and
${\cal A}_{CP}(K^+ \pi^0) = 0$ (dashed curves) or $|{\cal A}_{CP}(K^+ \pi^0)|
= 0.11$ (solid curve) as a function of the weak phase $\gamma$
\cite{GRKpi03}.
Horizontal dashed lines denote $\pm 1 \sigma$ experimental limits on $R_c$,
while dotdashed lines denote 95\% c.l. ($ \pm 1.96 \sigma$) limits.  We have
taken $\delta_{EW} = 0.80$ (its $1 \sigma$ upper limit), which
leads to the most conservative bound on $\gamma$.
\label{fig:Rcacp}}
\end{figure}

New data on branching ratios and CP asymmetries permit an update of previous
analyses \cite{GR02,NR}.  One makes use of the quantities (see \cite{CGR} and
\cite{GRKpi03})
\beq
R_c \equiv \frac{2 \gb(B^+ \to K^+ \pi^0)}{\gb(B^+ \to K^0 \pi^+)}
= 1.24 \pm 0.13~~, \label{eqn:Rc}
\eeq
\beq \label{eqn:Accp}
{\cal A}_{CP}(K^+ \pi^0)
 =  - \frac{2 r_c \sin \delta_c \sin \gamma}{R_c} = 0.035 \pm 0.071~~~,
\eeq
where $r_c \equiv |(T+C)/P| = 0.20 \pm 0.02$, and a strong phase $\delta_c$
is eliminated by combining (\ref{eqn:Rc}) and (\ref{eqn:Accp}).
One must also use an estimate \cite{NR} of the electroweak penguin parameter
$\delta_{\rm EW} = 0.65 \pm 0.15$.  One obtains the most conservative (i.e.,
weakest) bound on $\gamma$ for the maximum values of $r_c$ and $\delta_{\rm
EW}$ \cite{GR02}.  The resulting plot is shown in Fig.\ \ref{fig:Rcacp}
\cite{JRmor,GRKpi03}.) One
obtains a bound at the $1 \sigma$ level very similar to that in the previous
case:  $\gamma \gs 52^\circ$.  The bound is actually set by the curve for
{\it zero} CP asymmetry, as emphasized in Ref.\ \cite{NR}.

\subsection{Asymmetries in $B^+ \to (\pi^0,\eta,\eta')K^+$}
\begin{figure}
\includegraphics[height=4.2in]{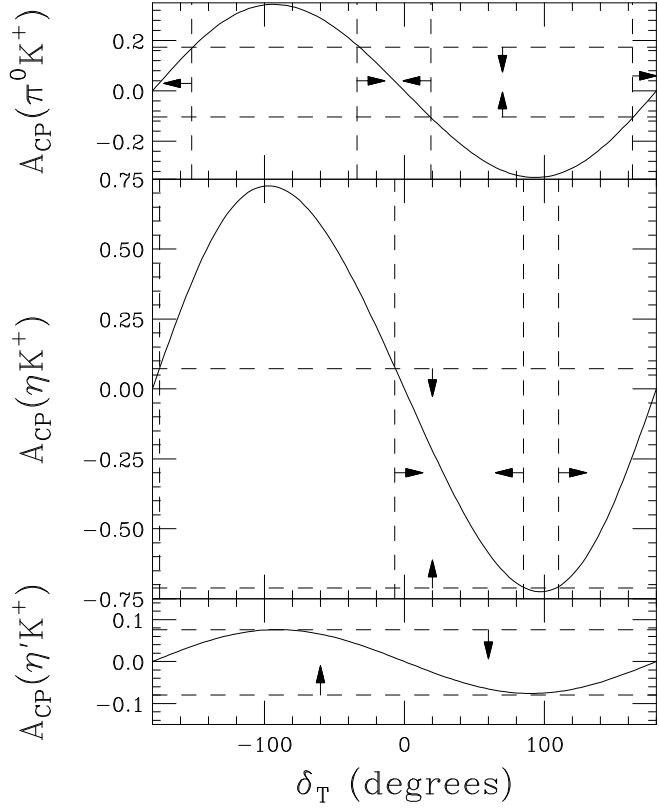}
\caption{Predicted $CP$ rate asymmetries when $\gamma = 60^\circ$ for $B^+ \to
\pi^0 K^+$ (top), $B^+ \to \eta K^+$ (middle), and $B^+ \to \eta' K^+$
(bottom) \cite{CGR}.  Horizontal dashed lines denote 95\% c.l. ($\pm 1.96
\sigma$) upper and lower experimental bounds, leading to corresponding bounds
on $\delta_T$ denoted by vertical dashed lines.  Arrows point toward allowed
regions.
\label{fig:acp}}
\end{figure}

The amplitudes for the decays $B^+ \to M^0 K^+$~ [($M^0 = (\pi^0,\eta,\eta')$]
all are dominated by penguin amplitudes and can be expressed as
\beq
A(B^+ \to M^0 K^+) = a(e^{i \gamma} - \delta_{EW}) e^{i \delta_T} - b~~~,
\eeq
where $a$ and $b$ may be calculated using flavor SU(3) from other processes
\cite{CGR}, and $\delta_T$ is a strong phase.  The allowed ranges of the
resulting CP asymmetries are shown in Fig.\ \ref{fig:acp} \cite{CGR}.  The
asymmetries are sensitive to $\delta_T$ but vary less significantly with
$\gamma$ over the 95\% c.l. allowed range \cite{CKMf} $38^\circ < \gamma <
80^\circ$.  For illustration we have chosen $\gamma = 60^\circ$.

The constraints on $\delta_T$ from $A_{CP}(\pi^0 K^+)$ are $-34^\circ
\le \delta_T \le 19^\circ$ and a region of comparable size around $\delta_T =
\pi$.  The allowed range of $A_{CP}(\eta K^+)$ restricts these regions
further, leading to net allowed regions $-7^\circ \le \delta_T \le 19^\circ$ or
a comparable region around $\delta_T = \pi$.  These regions do not change
much if we vary $\gamma$ over its allowed range.  The scheme of Ref.\ \cite{BN}
predicts an opposite sign of $A_{CP}(\eta K^+)$ to ours for a given
sign of $\delta_T$ and hence the constraints will differ.

\subsection{$B^+ \to \pi^+ \eta$}
The possibility that several different amplitudes could contribute to
$B^+ \to \pi^+ \eta$, thereby leading to the possibility of a large direct
CP asymmetry, has been recognized for some time \cite{GR95,DGR95,DGR97,BRS,AK}.
Contributions can arise from a tree amplitude (color-favored plus
color-suppressed) $T+C$, whose magnitude is estimated from that
occurring in $B^+ \to \pi^+ \pi^0$, a penguin amplitude $P$, obtained via
flavor SU(3) from $B^+ \to K^0 \pi^+$, and a singlet penguin amplitude $S$,
obtained from $B \to \eta' K$.

In Table \ref{tab:etapi} we summarize branching ratios and CP asymmetries
obtained for the decay $B^+ \to \pi^+ \eta$ by CLEO \cite{CLeta}, BaBar
\cite{Baeta}, and Belle \cite{Bebrs}.  We assume that the $S$ and $P$
amplitudes have the same weak and strong phases.  The equality of their weak
phases is quite likely, while tests exist for the latter assumption \cite{CGR}.

\begin{table}
\caption{Branching ratios and CP asymmetries for $B^+ \to \pi^+ \eta$.
\label{tab:etapi}}
\begin{tabular}{l c c} \hline
 & $\cB~(10^{-6})$ & $A_{CP}$ \\ \hline
CLEO \cite{CLeta}  & $1.2^{+2.8}_{-1.2}~(< 5.7)$ & -- \\
BaBar \cite{Baeta} & $4.2^{+1.0}_{-0.9} \pm 0.3$ & $-0.51^{+0.20}_{-0.18}$ \\
Belle \cite{Bebrs} & $5.2^{+2.0}_{-1.7} \pm 0.6$ & --  \\
Average            & $4.1 \pm 0.9$               & $-0.51^{+0.20}_{-0.18}$ \\
$|T+C|^2$ alone    &             3.5             &          0             \\
$|P+S|^2$ alone    &             1.9             &          0        \\
\hline
\end{tabular}
\end{table}

If an amplitude $A$ for a process receives two contributions with differing
strong and weak phases, one can write
\beq
A = a_1 + a_2 e^{i \phi} e^{i \delta}~~,~~~
\bar A = a_1 + a_2 e^{-i \phi} e^{i \delta}~~~.
\eeq
The CP-averaged decay rate is proportional to $a_1^2 + a_2^2 + 2 a_1 a_2
\cos \phi \cos \delta$, while the CP asymmetry is
\beq
A_{CP} = - \frac{2 a_1 a_2 \sin \phi \sin \delta}
{a_1^2 + a_2^2 + 2 a_1 a_2 \cos \phi \cos \delta}~~~.
\eeq
In the case of $B^+ \to \pi^+ \eta$ the rates and CP asymmetry suggest that
$|\sin \phi \sin \delta| > |\cos \phi \cos \delta|$.

By combining the branching ratio and $CP$ rate asymmetry information of the
$\pi^{\pm} \eta$ modes, one can extract the values of the relative
strong phase $\delta$ and the weak phase $\alpha$, assuming maximal 
constructive interference between ordinary and singlet penguin amplitudes.
On the basis of the range of amplitudes extracted from other processes, we find
that the rates and CP asymmetries for $B^+ \to \pi^\pm \eta$ and $B^+ \to
\pi^\pm \eta'$ are correlated with one another \cite{CGR}:
\beq
\label{eq:eq1}
A_{CP}(\pi^+ \eta) = -(0.91 \sin\delta \sin\alpha)/
(1 - 0.91 \cos\delta \cos\alpha) ~,
\eeq
\beq
\label{eq:eq2}
A_{CP}(\pi^+ \eta') = -(\sin\delta \sin\alpha)/
(1 - \cos\delta \cos\alpha) ~,
\eeq
\beq
\label{eq:eq3}
\cB(\pi^+ \eta) = 5 \times 10^{-6}(1 - 0.91 \cos\delta \cos\alpha) ~,
\eeq
\beq
\label{eq:eq4}
\cB(\pi^+ \eta') =  3.4 \times 10^{-6}(1 - \cos\delta \cos\alpha) ~,
\eeq
where $\cB$ refers to a CP-averaged branching ratio.  One finds that
$\cB(B^+ \to \pi^+ \eta') = (2.7 \pm 0.7) \times 10^{-6}$ (below current upper
bounds) and that $A_{CP}(\pi^+ \eta') = -0.57 \pm 0.23$.

\begin{figure}
\includegraphics[height=3.5in]{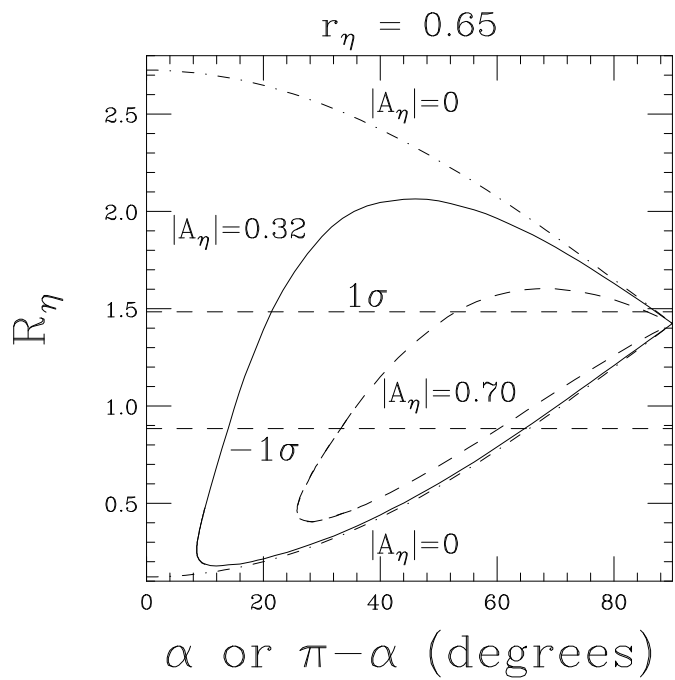}
\caption{Predicted value of $R_\eta$ (ratio of observed $CP$-averaged $B^\pm
\to \pi^\pm \eta$ decay rate to that predicted for tree amplitude alone) as a
function of $\alpha$ for various values of $CP$ asymmetry $|A_\eta|$
\cite{CGR}.  (The values 0.70 and 0.32 correspond to $\pm 1 \sigma$ errors on
this asymmetry.)
\label{fig:Reta}}
\end{figure}

The amplitudes for $B^\pm \to \pi^\pm \eta$ may be written in the form
\beq
A(\pi^\pm \eta) \sim e^{\pm i \gamma} \left[ 1 - r_\eta
e^{i (\pm \alpha + \delta)} \right]~~,\\
\eeq
where $r_\eta$ (estimated in Ref.\ \cite{CGR} to be $0.65 \pm 0.06$) is the
ratio of penguin to tree contributions to the $B^\pm \to \pi^\pm \eta$ decay
amplitudes.  We define $R_\eta$ as the ratio of the observed $CP$-averaged
$B^\pm \to \pi^\pm \eta$ decay rate to that which would be expected in the
limit of no penguin contributions and find
\beq
\label{eqn:Reta}
R_\eta = 1 + r_\eta^2 - 2 r_\eta \cos \alpha \cos \delta = 1.18 \pm 0.30~~.
\eeq
One can then use the information on the observed $CP$ asymmetry in this mode to
eliminate $\delta$ and constrain $\alpha$.  (For a related treatment with a
different convention for penguin amplitudes see Ref. \cite{MGFPCP}.)  The
asymmetry is
\beq
\label{eqn:Aeta}
A_\eta = -2 r_\eta \sin \alpha \sin \delta/R_\eta = -0.51 \pm 0.19~~,
\eeq
so one can either use the result
\beq
\label{eqn:RAeta}
R_\eta = 1 + r_\eta^2 \pm \sqrt{4 r_\eta^2 \cos^2 \alpha
- (A_\eta R_\eta)^2 \cot^2 \alpha}
\eeq
with experimental ranges of $R_\eta$ and $A_\eta$ or solve (\ref{eqn:RAeta})
for $R_\eta$ in terms of $\alpha$ and $A_\eta$.  The result of this latter
method is illustrated in Fig.\ \ref{fig:Reta} \cite{CGR}.

The range of $\alpha$ allowed at 95\% c.l.\ in standard-model fits to CKM
parameters is $78^\circ \le \alpha \le 122^\circ$ \cite{CKMf}.  For
comparison, Fig.\ \ref{fig:Reta} permits values of $\alpha$ in the three
ranges
\beq
14^\circ \le \alpha \le 53^\circ~~,~~~
60^\circ \le \alpha \le 120^\circ~~,~~~
127^\circ \le \alpha \le 166^\circ~~
\eeq
if $R_\eta$ and $|A_\eta|$ are constrained to lie within their $1 \sigma$
limits.  The middle range overlaps the standard-model parameters, restricting
them slightly.  Better constraints on $\alpha$ in this region would require
reduction of errors on $R_\eta$.

\section{Final-state phases}
\subsection{$B$ decays}
\begin{figure}
\includegraphics[height=2.1in]{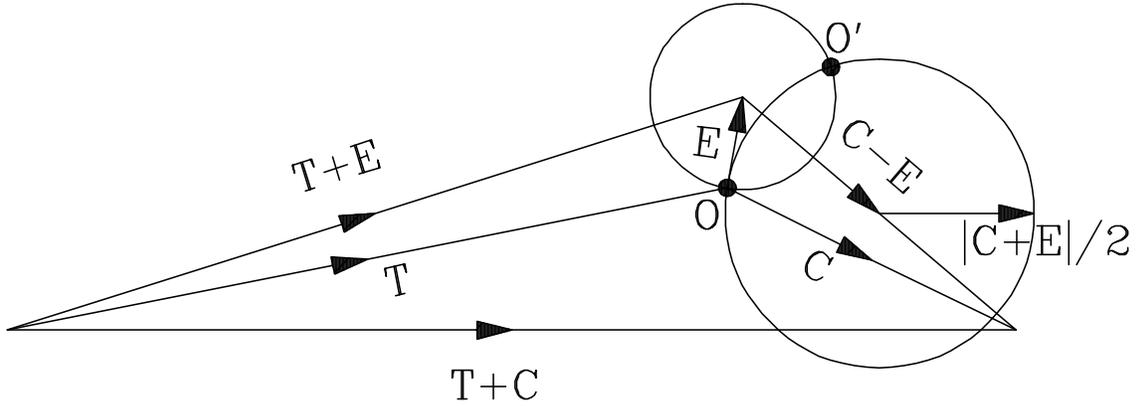}
\caption{Amplitude triangle for $\overline{B} \to D \pi$ and related decays
\cite{bdfsi}.
The amplitude $E$ points from either $O$ or $O'$ to the center of the small
circle.  The amplitudes $T$ and $C$ are shown only for the first of these two
solutions.  Here $A(B^0 \to D^- \pi^+) = T+E$, $A(B^+ \to \od \pi^+) = T+C$,
$\s A(B^0 \to \od \pi^0) = C-E$, $\st A(B^0 \to \od \eta) = -(C+E)$, and
$A(B^0 \to D_s^- K^+) = E$.
\label{fig:dpi}}
\end{figure}

We have seen that final-state phases are needed in order to observe direct
CP asymmetries.  It is interesting to obtain information on such phases in
those $D$ decays in which weak phases are expected to play little role,
so that magnitures of amplitudes directly reflect relative strong phases.
As one example we illustrate such phases in the decays of $B \to D \pi$ and
related processes in Fig.\ \ref{fig:dpi} \cite{bdfsi}.  The color-suppressed
amplitude $C$ is found to have a non-trivial strong phase with respect to
the color-favored tree amplitude $T$, with a small exchange amplitude $E$
(governing $B^0 \to D_s^- K^+$) at an even larger phase with respect to $T$.
Such large phases can signal strong rescattering effects.

\subsection{Charm decays}
In one method for measuring the weak phase $\gamma$ in $B^\pm \to K^\pm
(KK^*)_D$ decays, the relative strong phase $\delta_D$ in $D^0 \to K^{*+} K^-$
and $D^0 \to K^{*-} K^+$ decays (equivalently, in $D^0 \to K^{*+} K^-$ and $\od
\to K^{*+} K^-$) plays a role \cite{GLS}.  A study of the Dalitz plot in $D^0
\to K^+ K^- \pi^0$ can yield information on this phase \cite{RS03}.  By
comparing such Dalitz plots for constructive and destructive interference
between the two $K^*$ bands one finds that a clear-cut distinction is possible
between $\delta_D = 0$ and $\delta_D = \pm \pi$ with a couple of thousand
decays.

\section{$B_s$ mixing and decays}
\subsection{$B_s$--$\overline B_s$ mixing}
The ratio of the $B_s$--$\overline B_s$ mixing amplitude $\Delta m_s$ to the
$B^0$--$\ol B^0$ mixing amplitude $\Delta m_d$ ($B_d \equiv B^0$) is given by
\beq
\frac{\Delta m_s}{\Delta m_d} =
\frac{f_{B_s}^2 B_{B_s}}{f_{B_d}^2 B_{B_d}} \frac{m_{B_s}}{m_{B_d}}
\left| \frac{V_{ts}}{V_{td}} \right|^2 \simeq 48 \times 2^{\pm 1}~~~.
\eeq
Here $f_{B_{d,s}}$ are meson decay constants, while $B_{B_{d,s}}$ express the
degree to which the mixing amplitude is due to vacuum intermediate states.
A lattice
estimate of the ratio $\xi \equiv (f_{B_s}/f_{B_d})\sqrt{B_{B_s}/B_{B_d}}$
is $1.21 \pm 0.04 \pm 0.05$ \cite{Bec}.  We have taken
$|V_{td}| = A \lambda^3|1 - \bar \rho - i \bar \eta|=(0.8 \pm 0.2) A \lambda^3$
with $|V_{ts}| = A \lambda^2$ and $\lambda = 0.22$.  With \cite{LEPBOSC}
$\Delta m_d = 0.502 \pm 0.006$ ps$^{-1}$ one then predicts
$\Delta m_s = 24~{\rm ps}^{-1} \times 2^{\pm 1}$.
The lower portion of this range is already excluded by the bound \cite{LEPBOSC}
$\Delta m_s > 14.4~{\rm ps}^{-1}~(95\%~{\rm c.l.})$.
When $\Delta m_s$ is measured it will constrain $\bar \rho$ significantly.

\subsection{Decays to CP eigenstates}
\subsubsection{$B_s \to J/\psi \phi,~J/\psi \eta, \ldots$.}
Since the weak phase in $\bar b \to \bar c c \bar s$ is expected to be zero
while that of $B_s$--$\ol B_s$ mixing is expected to be very small,
one expects CP asymmetries to be only a few percent in
the standard model for those $B_s$ decays dominated by this quark subprocess.
The $B_s \to J/\psi \phi$ final state is not a CP eigenstate but the even and
odd CP components can be separated using angular analyses.  The final
states of $B_s \to J/\psi \eta$ and $B_s \to J/\psi \eta'$ are CP-even so no
such analysis is needed.

\subsubsection{$B_s \to K^+ K^-$ vs.\ $B^0 \to \pi^+ \pi^-$.}
A comparison of time-dependent asymmetries in $B_s \to K^+ K^-$ and $B^0 \to
\pi^+ \pi^-$ \cite{RFKK} allows one to separate out strong and weak phases
and relative tree and penguin contributions.  In $B_s \to K^+ K^-$ the $\bar b
\to \bar s$ penguin amplitude is dominant, while the strangeness-changing
tree amplitude $\bar b \to \bar u u \bar s$ is smaller.  In $B^0 \to \pi^+
\pi^-$ it is the other way around: The $\bar b \to \bar u u \bar d$ tree 
amplitude dominates, while the $\bar b \to \bar d$ penguin is
Cabibbo-suppressed.  The U-spin subgroup of SU(3), which interchanges $s$ and
$d$ quarks, relates each amplitude in one process to that in the other apart
from the CKM factors.

\subsubsection{$\overline B_s,~B^0 \to K^+ \pi^-$.}
In comparing $B_s \to K^+ K^-$ with $B^0 \to \pi^+ \pi^-$,
the mass peaks will overlap with one another if analyzed in terms of the same
final state (e.g., $\pi^+ \pi^-$) \cite{Jesik}.  Thus, in the absence of good
particle identification, a variant on this scheme employing the decays
$B^0 \to K^+ \pi^-$ and $B_s \to K^- \pi^+$ (also related to one another by
U-spin) may be useful \cite{GRKpi00}.  For these final states, kinematic
separation may be easier.  One can also study the time-dependence
of $B_s \to K^+ K^-$ while normalizing the penguin amplitude using $B_s \to
K^0 \ol K^0$ \cite{GRKK}.

\subsection{Other SU(3) relations}
The U-spin subgroup of SU(3) allows one to relate many other $B_s$ decays
besides those mentioned above to corresponding $B_d$ decays \cite{MGU}.
Particularly useful are relations between CP-violating rate {\it differences}.
One thus will have the opportunity to perform many tests of flavor SU(3) and
to learn a great deal more about final-state phase patterns when a variety of
$B_s$ decays can be studied.

\section{Excited states}
\subsection{Flavor tagging for neutral $B$ mesons}
A promising method for tagging the flavor of a neutral $B$ meson is to
study the charge of the leading light hadron accompanying the fragmentation
of the heavy quark \cite{AB,GNR,GRtag}.  For example, an initial $b$ will
fragment into a $\ol B^0$ by ``dressing'' itself with a $\bar d$.  The
accompanying $d$, if incorporated into a charged
pion, will end up in a $\pi^-$.  Thus a $\pi^-$ is more likely to be ``near'' a
$\ol B^0$ than to a $B^0$ in phase space.  This correlation
between $\pi^-$ and $\ol B^0$ (and the corresponding correlation between
$\pi^+$ and $B^0$) is also what one would expect on the basis of non-exotic
resonance formation.  Thus the study of the resonance spectrum of the excited
$B$ mesons which can decay to $B + \pi$ or $B^* + \pi$ is of special
interest \cite{EHQ}.  The lowest such mesons are the P-wave levels of a $\bar
b$ antiquark and a light ($u$ or $d$) quark.

\subsection{Excited $D_s$ states below $D^{(*)}K$ threshold}
In April of this year the BaBar Collaboration \cite{BaDs} reported a narrow
resonance at 2317 MeV decaying to $D_s \pi^0$.  This state was quickly
confirmed by CLEO \cite{CLDs}, who also presented evidence for a narrow state
at 2463 MeV decaying to $D_s^* \pi^0$.  Both states have been confirmed by
Belle \cite{BeDs}.  We mention briefly why these states came as surprises.

The previously known P-wave levels of a charmed quark $c$ and an antistrange
$\bar s$ were candidates for $J=1$ and $J=2$ states at 2535 and 2572 MeV
\cite{PDG}.  These levels have narrow widths behave as expected if the spin of
the $\bar s$ and the orbital angular momentum were coupled up to $j = 3/2$.
(One expects $j$-$j$ rather than $L$-$S$ coupling in a light-heavy system
\cite{DGG,JRPW,HQ}.)  If the $j=1/2$ states were fairly close to these in mass
one would then expect another $J=1$ state and a $J=0$ state somewhere above
2500 MeV.  Instead, the candidate for the $J=0$ $c \bar s$ state is the one at
2317 MeV, with the state at 2463 MeV the candidate for the second $J=1$ level.
Belle's observation of the decay $D_{sJ}(2463) \to D_s \gamma$ reinforces this
interpretation \cite{BeDs}.  Both states are narrow since they are too light to
decay respectively to $D K$ or $D^* K$.  They decay instead via
isospin-violating transitions.
They are either candidates for $D^{(*)} K$ molecules \cite{BCL}, or indications
of a broken chiral symmetry which places them as positive-parity partners of
the $D_s$ and $D_s^*$ negative-parity $c \bar s$ ground states \cite{BEH}.
Indeed, the mass splittings between the parity partners appear to be exactly as
predicted ten years ago \cite{BH}.  Potential-based quarkonium models have a
hard time accommodating such low masses \cite{CJ,SG,Col},

There should exist {\it non-strange} $j=1/2$ $0^+$ and $1^+$ states, lower in
mass than the $j=3/2$ states at 2422 and 2459 MeV \cite{PDG} but quite broad
since their respective $\ol D \pi$ and $\ol D^* \pi$ channels will be open.
The study of such states will be of great interest since the properties of the
corresponding $B$-flavored states will be useful in tagging the flavor of
neutral $B$ mesons.

\subsection{Narrow positive-parity states below $\ol B^{(*)} K$ threshold?}
If a strange antiquark can bind to a charmed quark in both negative- and
positive-parity states, the same must be true for a strange antiquark and
a $b$ quark.  One should then expect to see narrow $J^P = 0^+$ and $1^+$
states with the quantum numbers of $\ol B K$ and $\ol B^* K$ but below those
respective thresholds.  They should decay to $\ol B_s \pi^0$ and $\ol B_s^*
\pi^0$, respectively.  To see such decays one will need a multi-purpose
detector with good charged particle and $\pi^0$ identification!

\section{Exotic $Q=-1/3$ quarks}
Might there be heavier quarks visible at hadron colliders?  At present we
have evidence for three families of quarks and leptons belonging to
16-dimensional multiplets of the grand unified group SO(10) (counting
right-handed neutrinos as a reasonable explanation of the observed oscillations
between different flavors of neutrinos).  Just as SO(10) was pieced
together from multiplets of SU(5) with dimensions 1, 5, and 10, the smallest
representation of a still larger grand unified group could contain the
16-dimensional SO(10) spinor.  Such a group is E$_{\rm
6}$ \cite{GRS}.  Its smallest representation, of dimension 27, contains a
16-dimensional spinor, a 10-dimensional vector, and a singlet of SO(10).  The
10-dimensional vector contains vector-like isosinglet quarks ``$h$'' and
antiquarks $\bar h$ of charge
$Q = \pm 1/3$ and isodoublet leptons.  The SO(10) singlets are candidates for
sterile neutrinos, one for each family.

The new exotic $h$ quarks can mix with the $b$ quark and push its mass
down with respect to the top quark \cite{JRmix}.  Troy Andre and I
are looking at signatures of $h \bar h$
production in hadron colliders, to either set lower mass
limits or see such quarks through their decays to $Z + b$, $W + t$, and
possibly ${\rm Higgs} + b$.  The $Z$, for example, would be identified by its
decays to $\nu \bar \nu$, $\ell^+ \ell^-$, or jet $+$ jet, while the Higgs
boson would show up through its $b \bar b$ decay if it were far enough below
$W^+ W^-$ threshold.

\section{Summary}
The process $B^0 \to J/\psi K_S$ has provided spectacular confirmation of the
Kobayashi-Maskawa theory of CP violation, measuring $\beta$ to a few
degrees.  Now one is entering the territory of more difficult measurements.

The decay $B^0 \to \pi^+ \pi^-$ can give useful information on $\alpha$.  One
needs either a measurement of
$\cB(B^0 \to \pi^0 \pi^0)$ \cite{GL}, probably at the $10^{-6}$ level
(present limits \cite{Babrs,Bebrs,CLbrs} are several times that), or a
better estimate of the tree amplitude from $B \to \pi l \nu$ \cite{LR}.
Indeed, such an estimate has been presented recently \cite{LR03}.  The
BaBar and Belle experimental CP asymmetries \cite{Bapipi,Bepipi} will
eventually converge to one another, as did the initial measurements
of $\sin 2 \beta$ using $B^0 \to J/\psi K_S$.

The $B \to \phi K_S$ decay can display new physics via special $\bar b \to \bar
s s \bar s$ operators or effects on the $\bar b \to \bar s$ penguin.  Some
features of any new amplitude can be extracted from the data in a
model-independent way if one uses both rate and asymmetry information
\cite{CR03}.  While the effective value of $\sin 2 \beta$ in $B^0 \to \phi K_S$
seems to differ from its expected value by more than $2 \sigma$, CP asymmetries
in $B \to K_S (K^+ K^-)_{CP=+}$ do not seem anomalous.

The rate for $B \to \eta' K_S$ is not a problem for the standard model if one
allows for a modest flavor-singlet penguin contribution in addition to the
standard penguin amplitude.  The CP asymmetries for this process are in accord
with the expectations of the standard model at the $1 \sigma$ level or
better.  Effects of the singlet penguin amplitude may also be visible
elsewhere, for example in $B^+ \to p \bar p K^+$.

Various ratios of $B \to K \pi$ rates, when combined with information on
CP asymmetries, show promise for constraining phases in the CKM matrix.
These tests have steadily improved in accuracy in the past couple of years.
One expects further progress as $e^+ e^-$ luminosities increase, and as hadron
colliders begin to provide important contributions.  The decays $B^+ \to
\pi^+ \eta$ and $B^+ \to \pi^+ \eta'$ show promise for displaying large CP
asymmetries \cite{CGR} since they involve contributions of different amplitudes
with comparable magnitudes.  Strong final-state phases, important for the
observation of direct CP violation, are beginning to be mapped out in $B$
decays.

In the near term the prospects for learning about the $B_s$--$\ol B_s$ mixing
amplitude are good.  The potentialities of hadron colliders for the study
study of CP violation and branching ratios in $B_s$ decays will be
limited only by the versatility of detectors.  Surprises in spectroscopy,
as illustrated by the low-lying positive-parity $c \bar s$ candidiates, still
can occur, and one is sure to find more of them.
Finally, one can search for objects related to the properties of $b$ quarks,
such as the exotic isosinglet quarks $h$, with improved sensitivity
in Run II of the Tevatron and with greatly expanded reach at the LHC.

\begin{theacknowledgments}
I thank my collaborators on topics mentioned here:
Troy Andre, Cheng-Wei Chiang, Michael Gronau, Zumin Luo, and Denis Suprun.
Michael Gronau and Hassan Jawahery also made helpful comments on the manuscript.
This work was supported in part by the United States Department of Energy
under Grant No.\ DE FG02 90ER40560.
\end{theacknowledgments}

\def \ajp#1#2#3{Am.\ J. Phys.\ {\bf#1}, #2 (#3)}
\def \apny#1#2#3{Ann.\ Phys.\ (N.Y.) {\bf#1}, #2 (#3)}
\def \app#1#2#3{Acta Phys.\ Polonica {\bf#1}, #2 (#3)}
\def \arnps#1#2#3{Ann.\ Rev.\ Nucl.\ Part.\ Sci.\ {\bf#1}, #2 (#3)}
\def \art{and references therein}
\def \cmts#1#2#3{Comments on Nucl.\ Part.\ Phys.\ {\bf#1} (#3) #2}
\def \cn{Collaboration}
\def \cp89{{\it CP Violation,} edited by C. Jarlskog (World Scientific,
Singapore, 1989)}
\def \econf#1#2#3{Electronic Conference Proceedings {\bf#1}, #2 (#3)}
\def \epjc#1#2#3{Eur.\ Phys.\ J.\ C {\bf#1} (#3) #2}
\def \f79{{\it Proceedings of the 1979 International Symposium on Lepton and
Photon Interactions at High Energies,} Fermilab, August 23-29, 1979, ed. by
T. B. W. Kirk and H. D. I. Abarbanel (Fermi National Accelerator Laboratory,
Batavia, IL, 1979}
\def \hb87{{\it Proceeding of the 1987 International Symposium on Lepton and
Photon Interactions at High Energies,} Hamburg, 1987, ed. by W. Bartel
and R. R\"uckl (Nucl.\ Phys.\ B, Proc.\ Suppl., vol. 3) (North-Holland,
Amsterdam, 1988)}
\def \ib{{\it ibid.}~}
\def \ibj#1#2#3{~{\bf#1} (#3) #2}
\def \ichep72{{\it Proceedings of the XVI International Conference on High
Energy Physics}, Chicago and Batavia, Illinois, Sept. 6 -- 13, 1972,
edited by J. D. Jackson, A. Roberts, and R. Donaldson (Fermilab, Batavia,
IL, 1972)}
\def \ijmpa#1#2#3{Int.\ J.\ Mod.\ Phys.\ A {\bf#1} (#3) #2}
\def \ite{{\it et al.}}
\def \jhep#1#2#3{JHEP {\bf#1} (#3) #2}
\def \jpb#1#2#3{J.\ Phys.\ B {\bf#1}, #2 (#3)}
\def \lg{{\it Proceedings of the XIXth International Symposium on
Lepton and Photon Interactions,} Stanford, California, August 9--14, 1999,
edited by J. Jaros and M. Peskin (World Scientific, Singapore, 2000)}
\def \lkl87{{\it Selected Topics in Electroweak Interactions} (Proceedings of
the Second Lake Louise Institute on New Frontiers in Particle Physics, 15 --
21 February, 1987), edited by J. M. Cameron \ite~(World Scientific, Singapore,
1987)}
\def \kaon{{\it Kaon Physics}, edited by J. L. Rosner and B. Winstein,
University of Chicago Press, 2001}
\def \kdvs#1#2#3{{Kong.\ Danske Vid.\ Selsk., Matt-fys.\ Medd.} {\bf #1}, No.\
#2 (#3)}
\def \ky{{\it Proceedings of the International Symposium on Lepton and
Photon Interactions at High Energy,} Kyoto, Aug.~19-24, 1985, edited by M.
Konuma and K. Takahashi (Kyoto Univ., Kyoto, 1985)}
\def \mpla#1#2#3{Mod.\ Phys.\ Lett.\ A {\bf#1} (#3) #2}
\def \nat#1#2#3{Nature {\bf#1}, #2 (#3)}
\def \nc#1#2#3{Nuovo Cim.\ {\bf#1} (#3) #2}
\def \nima#1#2#3{Nucl.\ Instr.\ Meth.\ A {\bf#1}, #2 (#3)}
\def \np#1#2#3{Nucl.\ Phys.\ {\bf#1} (#3) #2}
\def \npps#1#2#3{Nucl.\ Phys.\ Proc.\ Suppl.\ {\bf#1} (#3) #2}
\def \npbps#1#2#3{Nucl.\ Phys.\ B Proc.\ Suppl.\ {\bf#1} (#3) #2}
\def \os{XXX International Conference on High Energy Physics, Osaka, Japan,
July 27 -- August 2, 2000}
\def \PDG{Particle Data Group, K. Hagiwara \ite, \prd{66}{010001}{2002}}
\def \pisma#1#2#3#4{Pis'ma Zh.\ Eksp.\ Teor.\ Fiz.\ {\bf#1}, #2 (#3) [JETP
Lett.\ {\bf#1}, #4 (#3)]}
\def \pl#1#2#3{Phys.\ Lett.\ {\bf#1} (#3) #2}
\def \pla#1#2#3{Phys.\ Lett.\ A {\bf#1}, #2 (#3)}
\def \plb#1#2#3{Phys.\ Lett.\ B {\bf#1} (#3) #2}
\def \pr#1#2#3{Phys.\ Rev.\ {\bf#1} (#3) #2}
\def \prc#1#2#3{Phys.\ Rev.\ C {\bf#1} (#3) #2}
\def \prd#1#2#3{Phys.\ Rev.\ D {\bf#1} (#3) #2}
\def \prl#1#2#3{Phys.\ Rev.\ Lett.\ {\bf#1} (#3) #2}
\def \prp#1#2#3{Phys.\ Rep.\ {\bf#1} (#3) #2}
\def \ptp#1#2#3{Prog.\ Theor.\ Phys.\ {\bf#1} (#3) #2}
\def \rmp#1#2#3{Rev.\ Mod.\ Phys.\ {\bf#1} (#3) #2}
\def \rp#1{~~~~~\ldots\ldots{\rm rp~}{#1}~~~~~}
\def \si90{25th International Conference on High Energy Physics, Singapore,
Aug. 2-8, 1990}
\def \slc87{{\it Proceedings of the Salt Lake City Meeting} (Division of
Particles and Fields, American Physical Society, Salt Lake City, Utah, 1987),
ed. by C. DeTar and J. S. Ball (World Scientific, Singapore, 1987)}
\def \slac89{{\it Proceedings of the XIVth International Symposium on
Lepton and Photon Interactions,} Stanford, California, 1989, edited by M.
Riordan (World Scientific, Singapore, 1990)}
\def \smass82{{\it Proceedings of the 1982 DPF Summer Study on Elementary
Particle Physics and Future Facilities}, Snowmass, Colorado, edited by R.
Donaldson, R. Gustafson, and F. Paige (World Scientific, Singapore, 1982)}
\def \smass90{{\it Research Directions for the Decade} (Proceedings of the
1990 Summer Study on High Energy Physics, June 25--July 13, Snowmass, Colorado),
edited by E. L. Berger (World Scientific, Singapore, 1992)}
\def \tasi{{\it Testing the Standard Model} (Proceedings of the 1990
Theoretical Advanced Study Institute in Elementary Particle Physics, Boulder,
Colorado, 3--27 June, 1990), edited by M. Cveti\v{c} and P. Langacker
(World Scientific, Singapore, 1991)}
\def \yaf#1#2#3#4{Yad.\ Fiz.\ {\bf#1}, #2 (#3) [Sov.\ J.\ Nucl.\ Phys.\
{\bf #1}, #4 (#3)]}
\def \zhetf#1#2#3#4#5#6{Zh.\ Eksp.\ Teor.\ Fiz.\ {\bf #1}, #2 (#3) [Sov.\
Phys.\ - JETP {\bf #4}, #5 (#6)]}
\def \zpc#1#2#3{Zeit.\ Phys.\ C {\bf#1}, #2 (#3)}
\def \zpd#1#2#3{Zeit.\ Phys.\ D {\bf#1}, #2 (#3)}


\begin{thebibliography}{99}

\bibitem{JRmor} J. L. Rosner, presented at 38th Rencontres de Moriond on
Electroweak Interactions and Unified Theories, Les Arcs, France, 15--22
March 2003, \efi 03-16, hep-ph/0304200, to be published in the Proceedings.

\bibitem{JRLHC} J. L. Rosner, Enrico Fermi Institute Report No.\ EFI-03-26,
hep-ph/0305315, invited talk at 4th International Symposium on LHC Physics and
Detectors (LHC 2003), Batavia, Illinois, 1--3 May 2003, to be published in
Eur.\ Phys.\ J.

\bibitem{WP} L. Wolfenstein, \prl{51}{1945}{1983}.

\bibitem{Battaglia}
M.~Battaglia {\it et al.}, to appear as a CERN Yellow Report, based on the
Workshop on CKM Unitarity Triangle (CERN 2002-2003), Geneva, Switzerland,
preprint hep-ph/0304132.

\bibitem{CKMf} A. H\"ocker \ite, \epjc{21}{225}{2001}.  Updated results
may be found on the web site {\tt http://ckmfitter.in2p3.fr/}.

\bibitem{BaBarPhys} {\it The BaBar Physics Book:  Physics at an Asymmetric
$B$ Factory}, edited by P. F. Harrison \ite, SLAC Report No.\ SLAC-R-0504,
1998.

\bibitem{TASI} J. L. Rosner, in {\it Flavor Physics for the Millennium}
(TASI 2000), edited by J. L. Rosner (World Scientific, Singapore, 2001),
p.\ 431.

\bibitem{Babeta} BaBar \cn, B. Aubert \ite, \prl{89}{201802}{2002}.

\bibitem{Bebeta} Belle \cn, K. Abe \ite, \prd{66}{071102}{2002}.

\bibitem{avbeta} Y. Nir, presented at XXXI International Conference
on High Energy Physics, Amsterdam, July, 2002, \npbps{117}{111}{2003}.

\bibitem{Ciu} M. Ciuchini \ite, \jhep{0107}{013}{2001}.

\bibitem{AL} A. Ali and D. London, \epjc{18}{665}{2001}.

\bibitem{GL} M. Gronau and D. London, \prl{65}{3381}{1990}.

\bibitem{GR02} M. Gronau and J. L. Rosner, \prd{65}{013004}{2002}; \ibj{65}
{079901(E)}{2002}; \ibj{65}{093012}{2002}.

\bibitem{GRconv} M. Gronau and J. L. Rosner, \prd{66}{053003}{2002};
\ibj{66}{119901(E)}{2002}.

\bibitem{SW} J. P. Silva and L. Wolfenstein, \prd{49}{1151}{1994}.

\bibitem{GHLR} M. Gronau, O. F. Hernandez, D. London, and J. L. Rosner,
\prd{50}{4529}{1994}; \ibj{52}{6356}{1995}; \ibj{52}{6374}{1995}.

\bibitem{Charles} J. Charles, \prd{59}{054007}{1999}.

\bibitem{GR95} M. Gronau and J. L. Rosner, \prd{53}{2516}{1996};
\prl{76}{1200}{1996}.

\bibitem{LR} Z. Luo and J. L. Rosner, \prd{65}{054027}{2002}.

\bibitem{Bapipi} BaBar \cn, B. Aubert et al., \prl{89}{281802}{2002}.

\bibitem{Bepipi} Belle \cn, K. Abe \ite, \prd{68}{012001}{2003}.

\bibitem{LR03} Z. Luo and J. L. Rosner, \efi 03-25, hep-ph/0305262,
to be published in Phys.\ Rev.\ D.

\bibitem{CLEOsl03} CLEO Collaboration, S. B. Athar \ite, Cornell University
Report No.\ CLNS 03/1819, hep-ex/0304019, submitted to Phys.\ Rev.\ D.

\bibitem{GW} Y. Grossman and M. Worah, \plb{395}{241}{1997}.

\bibitem{Baphks} G. Hamel de Monchenault, 38th Rencontres de
Moriond on Electroweak Interactions and Unified Theories \cite{JRmor},
hep-ex/0305055.

\bibitem{Bephks} Belle \cn, K. Abe \ite, \prd{67}{031102}{2003}.

\bibitem{Aubert:2003tk}
BaBar Collaboration, B.~Aubert \ite, SLAC-Report No.\ SLAC-PUB-9684,
hep-ex/0303029, 38th Rencontres de Moriond on Electroweak
Interactions and Unified Theories \cite{JRmor}.

\bibitem{npphks}
G.~Hiller, \prd{66}{071502}{2002};
M.~Ciuchini and L.~Silvestrini, \prl{89}{231802}{2002};
A.~Datta, \prd{66}{071702}{2002}; M.~Raidal, \prl{89}{231803}{2002};
B.~Dutta, C.~S.~Kim, and S.~Oh, \prl{90}{011801}{2003};
S.~Khalil and E.~Kou, \prd{67}{055009}{2003} and hep-ph/0303214;
G.~L.~Kane, P.~Ko, H.~Wang, C.~Kolda, J.~h.~Park and L.~T.~Wang,
\prl{90}{141803}{2003}; S.~Baek, \prd{67}{096004}{2003};
A. Kundu and T. Mitra, \prd{67}{116005}{2003};
K. Agashe and C. D. Carone, hep-ph/0304229.

\bibitem{CR03} C.-W. Chiang and J. L. Rosner, \prd{68}{014007}{2003}.

\bibitem{GLNQ} Y. Grossman, Z. Ligeti, Y. Nir, and H. Quinn, \prd{68}
{015004}{2003}.

\bibitem{GRKKK} M. Gronau and J. L. Rosner, \plb{564}{90}{2003}.

\bibitem{DGR95} A. S. Dighe, M. Gronau, and J. L. Rosner, \plb{367}{357}{1996};
\ibj{377}{325(E)}{1996}.

\bibitem{DGR97} A. S. Dighe, M. Gronau, and J. L. Rosner, \prl{79}{4333}{1997}.

\bibitem{CR01} C.-W. Chiang and J. L. Rosner, \prd{65}{074035}{2002}.

\bibitem{FHH} H.-K. Fu, X.-G. He, and Y.-K. Hsiao, preprint hep-ph/0304242.

\bibitem{CGR} C.-W. Chiang, M. Gronau, and J. L. Rosner, \efi 03-24,
hep-ph/0306021, to be published in Phys.\ Rev.\ D.

\bibitem{BN} M. Beneke and M. Neubert, \np{B651}{225}{2003}.

\bibitem{Kpp} Belle \cn, K. Abe \ite, \prl{88}{181803}{2002}.

\bibitem{JRbbbar} J. L. Rosner, \prd{68}{014004}{2003}.

\bibitem{FM} R. Fleischer and T. Mannel, \prd{57}{2752}{1998}.

\bibitem{Babrs} BaBar \cn, B. Aubert \ite, quoted by S. Playfer at LHCb
Workshop, CERN, February 2003; updated results on $B^+ \to K^0 \pi^+$ presented
by J. Ocariz at EPS 2003 (International Europhysics Conference on High
Energy Physics, Aachen, 17--23 July 2003.

\bibitem{Bebrs} Belle \cn, presented by T. Tomura at 38th Rencontres de Moriond
on Electroweak Interactions and Unified Theories \cite{JRmor}, hep-ex/0305036.

\bibitem{CLbrs} CLEO \cn, A. Bornheim \ite, Cornell Laboratory of Nuclear
Science Report No.\ CLNS-03-1816, hep-ex/0302026, submitted to Phys.\ Rev.\ D.

\bibitem{GRKpi03} M. Gronau and J. L. Rosner, hep-ph/0307095, submitted
to Phys.\ Lett.\ B.

\bibitem{LEPBOSC} LEP B Oscillations Working Group,
{\tt http://lepbosc.web.cern.ch/LEPBOSC/}.

\bibitem{GRKpi} M. Gronau and J. L. Rosner, \prd{57}{6843}{1998}.

\bibitem{MGFPCP} M. Gronau, presented at Flavor Physics and CP Violation
Conference, Paris, France, June 2003.

\bibitem{GRL} M. Gronau, J. L. Rosner, and D. London, \prl{73}{21}{1994}.

\bibitem{EWP} R. Fleischer, \plb{365}{399}{1994}; N. G. Deshpande and X.-G He,
\prl{74}{26}{1995}; \ibj{74}{4099(E)}{1995}.

\bibitem{NR} M. Neubert and J. L. Rosner, \plb{441}{403}{1998};
\prl{81}{5076}{1998}; M. Neubert, \jhep{9902}{014}{1999}.

\bibitem{BRS} S.~Barshay, D.~Rein and L.~M.~Sehgal, \plb{259}{475}{1991}.

\bibitem{AK} M. R. Ahmady and E. Kou, \prd{59}{054014}{1999}.

\bibitem{CLeta} CLEO Collaboration, S. J. Richichi \ite, \prl {85}{520}{2000}.

\bibitem{Baeta} BaBar Collaboration, B.~Aubert {\it et al.}, SLAC Report No.\
SLAC-PUB-9962, hep-ex/0303039, 38th Rencontres de Moriond on QCD
and High Energy Hadronic Interactions, 22--29 March, 2003, Les Arcs, France.

\bibitem{bdfsi} C.-W. Chiang and J. L. Rosner, \prd{67}{074013}{2003}.

\bibitem{GLS} Y. Grossman, Z. Ligeti, and A. Soffer, \prd{67}{071301}{2003}.

\bibitem{RS03} J. L. Rosner and D. A. Suprun, \efi 03-07, hep-ph/0303117,
to be published in Phys.\ Rev.\ D.

\bibitem{Bec} D. Becirevic,  38th Rencontres de Moriond on
Electroweak Interactions and Unified Theories \cite{JRmor}.

\bibitem{RFKK} R. Fleischer, \plb{459}{306}{1999}.

\bibitem{Jesik} R. Jesik and M. Pettini, in {\it $B$ Physics at the Tevatron:
Run II and Beyond}, Fermilab Report No.\ FERMILAB-Pub-01/197, hep-ph/0201071,
p.\ 179.  [See especially Fig.\ 6.12(b)].

\bibitem{GRKpi00} M. Gronau and J. L. Rosner, \plb{482}{71}{2000}.

\bibitem{GRKK} M. Gronau and J. L. Rosner, \prd{65}{113008}{2002}.

\bibitem{MGU} M. Gronau, \plb{492}{297}{2000}.

\bibitem{AB} A. Ali and F. Barreiro, \zpc{30}{635}{1986}.

\bibitem{GNR} M. Gronau, A. Nippe, and J. L. Rosner, \prd{47}{1988}{1993}.

\bibitem{GRtag} M. Gronau and J. L. Rosner, \prl{72}{195}{1994};
\prd{49}{254}{1994}; \ibj{63}{054006}{2001}; \ibj{64}{099902(E)}{2001}.

\bibitem{EHQ} E. Eichten, C. Hill, and C. Quigg, \prl{71}{4116}{1993}.

\bibitem{BaDs} BaBar \cn, B. Aubert \ite, \prl{90}{242001}{2003}.

\bibitem{CLDs} CLEO \cn, D. Besson \ite, Cornell University Report No.\
CLNS 03/1826, hep-ex/0305100, submitted to Phys.\ Rev.\ D.

\bibitem{BeDs} Belle \cn, presented by R. Chistov at Flavor Physics and CP
Violation Conference, Paris, France, June 2003;
Belle Collaboration, Belle Report BELLE-CONF-0334, hep-ex/0307041, July 2003.

\bibitem{PDG} \PDG.

\bibitem{DGG} A. De R\'ujula, H. Georgi, and S. L. Glashow,
\prl{37}{785}{1976}.

\bibitem{JRPW} J. L. Rosner, \cmts{16}{109}{1986}.

\bibitem{HQ} M. Lu, M. Wise, and N. Isgur, \prd{45}{1553}{1992}.
 
\bibitem{BCL} T. Barnes, F. Close, and H. J. Lipkin, preprint hep-ph/0305025.

\bibitem{BEH} W. A. Bardeen, E. Eichten, and C. T. Hill, Fermilab Report No.
FERMILAB-PUB-03-071-T, hep-ph/0305049.

\bibitem{BH} W. A. Bardeen and C. T. Hill, \prd{49}{409}{1994}.  Chiral
partners of the ground states of heavy mesons were independently predicted by
M. A. Nowak, M. Rho, and I. Zahed, \prd{48}{4370}{1993}.  See also D. Ebert, T.
Feldmann, R. Friedrich, and H. Reinhardt, \np{B434}{619}{1995}; D. Ebert,
T. Feldmann, and H. Reinhardt, \plb{388}{154}{1996}.

\bibitem{CJ} R. N. Cahn and J. D. Jackson, Lawrence Berkeley National
Laboratory Report No.\ LBNL-52572, hep-ph/0305012 (unpublished).

\bibitem{SG} S. Godfrey, preprint hep-ph/0305122 (unpublished).

\bibitem{Col} P. Colangelo and F. De Fazio, INFN Bari Report No.\
BARI-TH-03-462, hep-ph/0305140 (unpublished).

\bibitem{GRS} F. G\"ursey, P. Ramond, and P. Sikivie, \plb{60}{177}{1976}.

\bibitem{JRmix} J. L. Rosner, \prd{61}{097303}{2000}.

\end{thebibliography}
\end{document}